\documentclass[aps,prb,twocolumn,superscriptaddress,floatfix,longbibliography]{revtex4-2}

\usepackage{graphicx,graphics}
\usepackage{dcolumn}
\usepackage{amsmath}
\usepackage{amsfonts}
\usepackage{amssymb}
\usepackage{latexsym,verbatim}
\usepackage{physics}
\usepackage{bm}
\usepackage{color}
\usepackage{xcolor}
\usepackage[normalem]{ulem}
\usepackage[percent]{overpic}
\usepackage{bbold}
\usepackage{lipsum}
\usepackage{ragged2e}
\usepackage[breaklinks=true,colorlinks,citecolor=blue,linkcolor=blue,urlcolor=blue]{hyperref}

\usepackage{graphicx}

\DeclareMathAlphabet\mathbfcal{OMS}{cmsy}{b}{n}

\newcommand{\bp}{{\bm p}}
\newcommand{\bq}{{\bm q}}
\newcommand{\br}{{\bm r}}
\newcommand{\bu}{{\bm u}}

\newcommand{\vD}{v_{\mathrm{D}}}

\newcommand{\epsF}{\varepsilon_{\rm F}}

\usepackage{soul}
\usepackage{orcidlink}

\begin{document}

\title{Localized surface plasmons in a Weyl semimetal nanosphere}

\author{Francesco M. D.~Pellegrino~\orcidlink{0000-0001-5425-1292}}
\affiliation{Dipartimento di Fisica e Astronomia ``Ettore Majorana'',  Universit\`a di Catania, via S.\ Sofia, 64, I-95123 Catania,~Italy}
\affiliation{INFN, Sez.\ Catania, I-95123 Catania,~Italy}
%
\author{Francesco Buccheri~\orcidlink{0000-0002-9315-6199}}
\affiliation{Dipartimento Scienza Applicata e Tecnologia, Politecnico di Torino,
Corso Duca degli Abruzzi 24, I-10129, Torino,~Italy}
\affiliation{INFN Sezione di Torino, Via P. Giuria 1, I-10125, Torino,~Italy}
\author{G. G. N.~Angilella~\orcidlink{0000-0003-2687-9503}}
\affiliation{Dipartimento di Fisica e Astronomia ``Ettore Majorana'',  Universit\`a di Catania, via S.\ Sofia, 64, I-95123 Catania,~Italy}
\affiliation{INFN, Sez.\ Catania, I-95123 Catania,~Italy}
\affiliation{Scuola Superiore di Catania, Universit\`a di Catania, 9, via Valdisavoia, I-95123 Catania, Italy}
\affiliation{Centro Siciliano di Fisica Nucleare e Struttura della Materia, Catania, Italy}

\begin{abstract}
In this study, we investigate the localized surface plasmon modes of a sub-wavelength spherical nanoparticle composed of a Weyl semimetal, taking into account the axion modification of electrodynamics.
We derive analytical solutions for dipole and quadrupole normal modes by employing the quasistatic approximation. The axion term leads to modified Fr\"ohlich conditions, resulting in multiple non-degenerate plasmonic resonances with distinct polarization dependencies.
In contrast to isotropic conventional metals, the magnetoelectric properties of Weyl semimetals enable an incident electromagnetic field, with the electric field transverse to the surface of the sphere, to excite a localized surface plasmon.
\end{abstract}

\maketitle

\section{Introduction}

Localized surface plasmons (LSPs) are collective oscillations of conduction electrons excited by light on the surfaces of metal nanoparticles~\cite{maier_book}.
There has been a remarkable interest in LSPs due to their properties, such as confining light to nanoscale dimensions~\cite{bozhevolnyi_nature_2006}, allowing the manipulation of light energy and its spectral characteristics~\cite{li_acsnano_2014}, as well as maintaining phase coherence~\cite{fakonas_natph_2014}, and quantum coherence~\cite{chang_prl_2006,ridolfo_prl_2010}.
Traditionally, LSPs have been studied in noble metal nanostructures described by the Drude model, where the dielectric response is isotropic and simple resonance conditions govern the plasmonic spectrum.
Metal nanostructures have been utilized as nanoscale heat sources that can be tuned using light, leading to the development of thermoplasmonics~\cite{baffou_natmat_2020}. Moreover, LSPs play a central role in modern analytic tools \cite{stewart_cr_2008}, e.g., optical biosensors \cite{nanda_jpa_2024}.

Recently, Dirac nodal-line semimetals, such as NiSe and CoSe, with their anisotropic dielectric characteristics, have demonstrated the ability to support multiple localized surface plasmons within the optical range. This leads to a broad match with the spectrum of sunlight radiation, making them an excellent platform for thermoplasmonics~\cite{politano_small_2022,santoro_chem_2022,politano_advmat_2017,santoro_jmembrane_2022,santoro_jmembrane_2023}.
In the realm of Dirac materials~\cite{lupi_rev_2020}, Weyl semimetals (WSMs) represent 3D topological conductors characterized by unique electromagnetic properties and Fermi arc surface states~\cite{huang_prx_2015,lv_natphys_2015,xu_science_2015,Armitage_2018,ma_nature_2019,bonasera_prb_2022}. 
 The low-energy effective field theory of WSMs includes an axion-term in the electromagnetic Lagrangian, leading to modified Maxwell equations~\cite{wilczek_prl_1987,chen_prb_2013,zyuzin_prb_2015,kotov_prb_2018}. This axion term significantly modifies the helicon dispersion compared to traditional metals~\cite{pellegrino_prb_2015}, impacts the bulk and surface plasmons of extended samples~\cite{zhou_prb_2015, Andolina_2018_a, song_prb_2017, asgari_prb_2021,tsuchikawa_prb_2020,tamaya_jpcd_2019,bugaiko_prb_2020}, and the surface plasmon polaritons associated with cylindrical waveguides~\cite{peluso_2024}.
Although the study of collective excitations in the extended geometry of WSMs has been addressed~\cite{guo_light_2023}, the effect of the axion term on localized plasmon modes within finite systems is not as extensively investigated. 
This becomes especially relevant in thermoplasmonics and nanophotonics, where manipulating near-field resonances at the nanoscale is crucial. 
%
%
Here, we investigate the role of the axion electrodynamics in shaping the spectrum and field distribution of localized surface plasmons in a sub-wavelength WSM sphere. Using the quasistatic approximation, since retardation effects can be disregarded for sub-wavelength particles, we solve the modified Maxwell equations and derive the conditions for dipole and quadrupole normal modes~\cite{frohlich_book,bohren_book,maier_book,davis_rmp_2017}. We demonstrate that the axion-induced gyrotropic term not only modifies the Fr\"ohlich resonance condition but also introduces a dependence on the polarization of the incident field, leading to multiple non-degenerate resonance frequencies. These results mark a departure from the isotropic behavior of conventional plasmonic systems and reveal new opportunities for harnessing topological materials in nanoscale light–matter interaction platforms.


The paper is structured as follows. In Sect.~\ref{sec:model}, we introduce the theoretical framework based on axion-modified electrodynamics.  In Sect.~\ref{sec:electrostatics}, we solve the electrostatic problem of a WSM sphere placed in a uniform static field.  In Sect.~\ref{sec:quasi} we analyze dipole LSPs and derive the corresponding modified Fr\"ohlich condition. Sect.~\ref{sec:electrostatics-q} extends the electrostatic analysis of a WSM sphere placed in an electric field with quadrupole symmetry, and in Sect.~\ref{sec:quasi-q}. We analyze quadrupole LSPs and obtain the corresponding modified Fr\"ohlich condition. Sect.~\ref{sec:TE} explores the possibility of exciting LSPs by an electric field that is perpendicular to the surface of the sphere. Finally, conclusions are drawn in Section~\ref{sec:conclusions}.

\section{Axion-Maxwell equations}
\label{sec:model}


We consider the simplest model for a WSM~\cite{burkov_prl_2011,Burkov_2011_b,Armitage_2018}, where the Hamiltonian presents two nodes only
\begin{equation}\label{eq:WeylHamiltonian}
{\cal H} =   \hbar  v_{\rm D}\tau^z {\bm \sigma} \cdot\left(-i \nabla +\tau^z  {\bm b}\right) +\hbar \tau^z b_0~.
\end{equation}
Here, $v_{\rm D}$ is the Dirac-Weyl velocity, $\tau^z$ describes the node degree of freedom with chirality $\pm 1$, and the 3D vector of Pauli matrices $\bm \sigma=(\sigma^x,\sigma^y,\sigma^z)^\top$ describes conduction- and valence-band degrees of freedom. The two Weyl nodes are placed at $\pm {\bm b}$ and shifted by $2 \hbar b_0$ in energy. 

In Eq.~(\ref{eq:WeylHamiltonian}), the terms related to $b_0$ and the vector ${\bm b} = (b_x,b_y,b_z)^\top$ can be effectively removed by a unitary transformation~\cite{zyuzin_prb_2012}. After this transformation, the Hamiltonian simplifies to ${\cal H} = -i \hbar v_{\rm D}\tau^z {\bm \sigma} \cdot \nabla$. Nevertheless, the chiral anomaly~\cite{peskin} results in this gauge transformation introducing an extra term to the electromagnetic Lagrangian ${\cal L}_{\rm em}$, which characterizes the interaction between electromagnetic fields and 3D WSMs~\cite{zyuzin_prb_2012}:
\begin{equation}\label{eq:emLagrangian}
{\cal L}_{\rm em}=\frac{1}{8 \pi}
(\bm{\mathcal{E}}^2-\bm{\mathcal{B}}^2)-\rho \phi + \bm{\mathcal{J}} \cdot {\bm A} +  {\cal L}_{\theta}~,
\end{equation}
where
\begin{equation}\label{eq:axionterm}
 {\cal L}_{\theta} = -\frac{\alpha}{4 \pi^2}  \theta (\br, t) \bm{\mathcal{E}}\cdot \bm{\mathcal{B}}~.
\end{equation}
Here, $\alpha=e^2/(\hbar c) \simeq 1/137$ is the QED fine-structure constant
and $\theta (\br, t)\equiv 2 ({\bm b} \cdot \br -  b_0  t)$ is called the axion angle.
The axion term ${\cal L}_{ \theta}$ 
modifies two of the four Maxwell equations, namely those involving the electromagnetic sources, i.e.~\cite{wilczek_prl_1987}
\begin{equation}\label{eq:nablaE}
 \nabla \cdot \bm{\mathcal{E}} = 4 \pi \left(\rho + \frac{\alpha}{2 \pi^2} {\bm b} \cdot \bm{\mathcal{B}}\right)~,
\end{equation}
and
\begin{equation}\label{eq:nablaB}
-\frac{1}{c} \frac{\partial \bm{\mathcal{E}}}{\partial t} + \nabla \times \bm{\mathcal{B}}=  \frac{4 \pi}{c} \left(  {\bm J} 
-\frac{\alpha}{2 \pi^2}  {c \bm b} \times \bm{\mathcal{E}} + \frac{\alpha}{2 \pi^2}   b_0 \bm{\mathcal{B}}
\right)~.
\end{equation}
Faraday's law, given by $\nabla \times \bm{\mathcal{E}} = - c^{-1} \frac{\partial \bm{\mathcal{B}}}{\partial t}$, together with the equation $\nabla \cdot \bm{\mathcal{B}} = 0$ indicating the absence of free magnetic monopoles, stay unchanged.
In this work, we focus on the case with $b_0=0$, such that the inversion symmetry is conserved. We consider a finite $\bm{b}$ related to the breaking of the time-reversal symmetry, and without loss of generality, we express it as $\bm{b}=b {\bm e}_z$, with $b>0$.  
This phase has been predicted in different materials, such as  HgCr$_2$ Se$_4$~\cite{xu_prl_2011}, MnBi$_2$Te$_4$ ~\cite{li_scienceadv_2019}, MnSn$_2$Sb$_2$Te$_6$~\cite{gao_prb_2023,
boulton_jpcm_2024}, K$_2$Mn$_3$ (AsO$_4$  )$_3$, XCrTe, (X=K, Rb) ~\cite{liu_prb_2024}, Eu$_2$Ir$_2$O$_7$~\cite{sushkov_prb_2015}, 
Mn$_3$Sn~\cite{cao_prb_2023}, Co$_3$Sn$_2$S$_2$~\cite{lohani_prb_2023}.
%
By working in the harmonic time decomposition, one has $\bm{\mathcal{E}}=\bm E e^{-i \omega t}$, $\bm{\mathcal{B}}=\bm B e^{-i \omega t}$, and $\bm{\mathcal{J}}=\bm J e^{-i \omega t}$. 
We assume that the current density $\bm J=\sigma(\omega) \bm E$ is due to the contributions of the bound-charge contribution and of free bulk electrons~\cite{Grosso_2000}, $\sigma(\omega)=\sigma_{\rm b}(\omega)+\sigma_{\rm e}(\omega)$. 
The bound-charge contribution is effectively expressed as
$\sigma_{\rm b}(\omega)=-i(\epsilon_{\rm b}-1)\omega/(4 \pi)$, which introduces a simple dielectric constant $\epsilon_{\rm b}$.
Moreover, the contribution of the free-electron bulk is described by the optical conductivity of a WSM in the long-wavelength limit. At zero temperature, it is expressed as
\begin{subequations}
\begin{align}
\sigma_{\rm e}(\omega)&=\sigma_{\rm intra}(\omega)+\sigma_{\rm inter}(\omega)~,\\
\sigma_{\rm intra}(\omega)~&=\frac{n_{\rm e} e^2}{m_{\rm c}} \frac{i}{\omega+i \gamma}~,
\\
\sigma_{\rm inter}(\omega)~&=\frac{\alpha c \omega}{12 \pi v_{\rm D}}\Big[ \Theta(\hbar \omega-2 \epsF)-\frac{i}{\pi}\log\Big| \frac{4 \Lambda^2}{4 \epsF^2-\hbar^2\omega^2}\Big|\Big]~.
\end{align}
\end{subequations}
Here, the intraband term $\sigma_{\rm intra}(\omega)$ has the form of Drude optical conductivity~\cite{Grosso_2000}, where $m_{\rm c}=\epsF/\vD^2$ is the WSM cyclotron mass, $\epsF$ is the Fermi energy,  $n_{\rm e}=\epsF^3/(3 \pi^2 \hbar^3 \vD^3)$ is the electron density, and $\gamma$ is an effective damping rate, that in the ultraclean limit is replaced by $\gamma\to 0^+$. Meanwhile, the interband term $\sigma_{\rm inter}(\omega)$ includes an ultraviolet cutoff denoted by $\Lambda$~\cite{lv_intjmod_2013,zhou_prb_2015,hoffman_prb_2015,kotov_prb_2016,kotov_prb_2018}.
The polarization vector is defined as follows:
\begin{equation}
\bm P = \frac{1}{i \omega}\left[\sigma(\omega) - \frac{\alpha c}{2 \pi^2} \bm b \times \right] \bm E~.
\end{equation}
Besides accounting for both the bound charge density and the charge density of free bulk electrons, we include the axion term in the polarization vector,  which encodes the
anomalous Hall effect~\cite{vazifeh_prl_2013}.
Moreover, introducing the electric displacement field as $\bm D=\bm E+4 \pi\bm P$
allows it to be represented as ${\bm D} = \epsilon \cdot {\bm E}$ in terms of the permittivity tensor. This tensor is non-reciprocal and takes the form of a skew-symmetric matrix
\begin{equation}\label{eq:tensor}
\epsilon_{j k}=
\begin{bmatrix}
\epsilon_\parallel&-i\epsilon_\bot&0\\
i \epsilon_\bot&\epsilon_\parallel&0\\
0&0&\epsilon_\parallel&\\
\end{bmatrix}_{j k},
\end{equation}
where 
\begin{subequations}
\begin{align}
\epsilon_\parallel&=\epsilon_{\rm b}-\frac{\omega_{\rm pl}^2}{\omega(\omega + i\gamma)}
-\frac{4 \pi}{i \omega} \sigma_{\rm inter}(\omega)~, \label{eq:epspar}\\
\epsilon_\bot&=\frac{2 \alpha c}{\pi \omega} b~,\label{eq:epsbot}
\end{align}
\end{subequations}
and $\omega_{\rm pl}=\sqrt{4 \pi  e^2 n_{\rm e}/m_{\rm c}}$ is the plasma frequency.
The permittivity tensor can be compactly represented as
\begin{equation}\label{eq:eps-eig}
\epsilon = \epsilon_\parallel \bm{e}_z \otimes\bm{e_z} + (\epsilon_\parallel + \epsilon_\bot) \bm{e}_+ \otimes\bm{e_+}^\star + (\epsilon_\parallel - \epsilon_\bot) \bm{e}_- \otimes\bm{e_-}^\star,
\end{equation}
where the unit vectors $\bm{e}_z$ and $\bm{e}_\pm = (\bm{e}_x \pm i \bm{e}_y)/\sqrt{2}$ represent its eigenvectors.
The off-diagonal term $\epsilon_\bot$ formally resembles the one generated in a traditional metal by the application of an external magnetic field. Within the Drude description~\cite{Ziman_1972, pellegrino_prb_2015,chiriaco_prb_2018}, in the presence of a weak magnetic field applied along the $z$-axis, the off-diagonal component of the permittivity tensor of a conventional electron gas acquires a contribution of the form $\epsilon_{\bot} \approx \omega_c \omega_{\rm pl}^2/\omega^3$, valid in the limit of small cyclotron frequency, $\omega_{\rm c} \ll \omega$. In contrast, in a WSM, the intrinsic anomalous Hall effect yields a qualitatively similar structure, but with a distinct frequency dependence, scaling as $1/\omega$, rather than $1/\omega^3$.

By resorting to the continuity equation, $\partial_t \rho + \nabla \cdot \bm{\mathcal{J}}=0$, 
Maxwell Eqs.~(\ref{eq:nablaE}--\ref{eq:nablaB}) assume the usual forms in a medium
\begin{equation}\label{eq:nablaD}
\nabla \cdot \bm D=0~,
\end{equation}
\begin{equation}
-\frac{i \omega}{c} \bm D=\nabla \times \bm B~.
\end{equation}
Using the Maxwell equations in the medium, we derive a set of modified wave equations:
\begin{equation}
\Big( \frac{\epsilon_\parallel\omega^2}{c^2}+\nabla^2\Big) {\bm D}=-\nabla\times \nabla\times \Big[\big(1-\epsilon_\parallel\epsilon^{-1}\cdot\big) {\bm D}\Big]~,
\end{equation}
\begin{equation}
\Big( \frac{\epsilon_\parallel\omega^2}{c^2}+\nabla^2\Big) {\bm B}=-\nabla\times\Big[\big(1-\epsilon_\parallel\epsilon^{-1}\cdot) \nabla\times {\bm B} \Big]~,
\end{equation}
where the terms on the right-hand sides account for the modifications introduced by the axion term, and $\epsilon$ is the permittivity tensor defined in Eq.~\eqref{eq:tensor}. This correction is a characteristic gyrotropic term of magnetic materials, which is typically treated perturbatively, such as for studying how the magnetism of small magnetic spheres affects Mie scattering~\cite{tarento_pre_2004}. 

\section{A WSM sphere in a uniform electric field}
\label{sec:electrostatics}

We consider a WSM sphere of radius $R$ located at the origin, in the presence of a uniform and static electric field 
 oriented in a generic direction 
\begin{equation}\label{eq:E0}
\bm{E}_{0}=E_0 {\bm u}~,
\end{equation}
 with $\bm{u}=\cos(\beta) \bm{e}_z+\sin(\beta) \cos(\varphi)\bm{e}_x+\sin(\beta) \sin(\varphi)\bm{e}_y$. 
From here on out, we assume that the surrounding medium is isotropic
and non-absorbing with a dielectric constant $\epsilon_{\rm M}=\Re(\epsilon_{\rm M})>0$, and the material of the sphere has a permittivity tensor with the non-diagonal form expressed in Eq.~\eqref{eq:tensor}.
We write the electric displacement field as $\bm{D}=-\nabla F(\br)$.
Following Eq.~\eqref{eq:nablaD}, the potential $F(\br)$ solves 
Laplace equation $\nabla^2 F(\br)=0$, both inside and outside the sphere.
We decompose the potential in spherical harmonics
\begin{equation}
F(\br)=\sum_{\ell=0}^\infty \sum_{m=-\ell}^\ell F_{\ell m}(r) Y_{\ell m}(\theta,\phi)~,
\end{equation}
where the radial functions are defined in a piecewise manner as
\begin{equation}
F_{\ell m}(\bm{r})=f_{\ell m} (r)\Theta(R-r)+g_{\ell m} (r) \Theta(r-R)~,
\end{equation}
where $f_{\ell m}(r) =w_{\ell m} r^\ell$, $g_{\ell m}(r)=s_{\ell m} r^\ell + t_{\ell m} r^{-(\ell+1)}$, and $\Theta(r)$ denotes the Heaviside step function.
By imposing that the potential is finite at the origin, and tends to  $F(\br) \to - \epsilon_{\rm M}E_0 \bm{u}\cdot \br$  for $r \to \infty$,  then $s_{\ell m}=\delta_{\ell,1} S_m$
where $S_{m}=-\epsilon_{\rm M}{ E}_0 \sqrt{4\pi/3}[\delta_{m,0}\cos(\beta)-\delta_{m,+1}\sin(\beta)e^{-i \varphi}/\sqrt{2}+ \delta_{m,-1}\sin(\beta)e^{i \varphi}/\sqrt{2}  ]$.
The coefficients $w_{\ell m}$ and $t_{\ell m}$ are set by the boundary conditions on the surface of the sphere,
\begin{equation}\label{eq:bcE}
\bm{E}\times \bm{e}_r|_{r=R-\eta}=\bm{E}\times \bm{e}_r|_{r=R+\eta}
\end{equation}
\begin{equation}\label{eq:bcD}
\bm{D}\cdot\bm{e}_r|_{r=R-\eta}=\bm{D}\cdot \bm{e}_r|_{r=R+\eta}
\end{equation}
where $\bm{e}_r=\bm{r}/r$ denotes the unit vector directed perpendicularly towards the surface of the sphere, and $\eta=0^+$. 
The electric displacement field is explicitly expressed as
\begin{equation}
\begin{aligned}
\bm{D}&=-\sum_{\ell=0}^\infty \sum_{m=-\ell}^\ell \Big[\partial_r F_{\ell m}(r) \bm{Y}_{\ell m}(\theta,\phi)\\
&+\frac{F_{\ell m}(r)}{r}  \bm{\Psi}_{\ell m}(\theta,\phi)\Big]~,
\end{aligned}
\end{equation}
where $\bm{Y}_{\ell m}(\theta,\phi)=Y_{\ell m}(\theta, \phi) \bm{e}_r $, $\bm{\Psi}_{\ell m}(\theta,\phi)=r \nabla Y_{\ell m}(\theta, \phi)$ and $\bm{\Phi}_{\ell m}(\theta,\phi)=\bm{e}_r \times  \bm{\Psi}_{\ell m}(\theta, \phi)$ are the vector spherical harmonics, and represent an extension of the scalar spherical harmonics $Y_{\ell m}(\theta, \phi)$~\cite{barrera_epj_1985}. 
From Eq.~\eqref{eq:bcD}, one has
\begin{equation}
\partial_r f_{\ell m}(r)|_{r=R}=\partial_r g_{\ell m}(r)|_{r=R}~,
\end{equation}
which is written explicitly as 
\begin{equation}\label{eq:bcD2}
w_{\ell m}=s_{\ell m}-\frac{\ell+1}{\ell} \frac{t_{\ell m}}{R^{2\ell+1}}~.
\end{equation}
Because of the non-diagonal form of the permittivity tensor of the sphere, in general, the boundary condition in Eq.~\eqref{eq:bcE} can involve different spherical harmonics.
%
%
%
Following the established solution for the conventional case where the axial term is absent, that is, when ${\bm b}={\bm 0}$, we employ Ansatz $w_{\ell m}=\delta_{\ell, 1} W_m$ and $t_{\ell m}=\delta_{\ell, 1} T_m$. Consequently, from Eq.~\eqref{eq:bcE}, we derive
\begin{equation}
\frac{\epsilon_{\rm M}}{\epsilon_\parallel+m\epsilon_\bot}W_m= S_m+T_m/R^3~,
\end{equation}
and from Eq.~\eqref{eq:bcD}, one has
\begin{equation}
W_{m}=S_{ m}-2 \frac{T_{ m}}{R^{3}}~,
\end{equation}
where $m \in \{0,\pm 1\}$. The coefficients $W_m$ and $T_m$, are explicitly expressed as
\begin{subequations}\label{eq:WT}
\begin{align}
W_m&=\frac{3(\epsilon_\parallel+m\epsilon_\bot)}{2 \epsilon_{\rm M}+\epsilon_\parallel+m\epsilon_\bot} S_m~,\\
T_m&=\frac{\epsilon_{\rm M}-(\epsilon_\parallel+m\epsilon_\bot)}{2 \epsilon_{\rm M}+\epsilon_\parallel+m\epsilon_\bot} R^3 S_m~.
\end{align}
\end{subequations}
%
The derived solution illustrates that outside the sphere, 
to the externally applied electric field $\bm{E}_{0}$
it is added the electric field $\bm{E}_{\rm dip}$, which is given by 
\begin{equation}
\bm{E}_{\rm dip} =  \frac{3 \bm{e}_r (\bm{e}_r\cdot \bp)-\bp}{\epsilon_{\rm M} r^3}~,
\end{equation} 
which is the expression of a static electric field generated by a dipole at the origin. 
Here, this dipole moment is in general not oriented in the $\bm{u}$ direction, and is expressed as
\begin{equation}\label{eq:p}
\begin{aligned}
\bp&=\sum_{m=-1}^1 T_m \bm{\Psi}_{1m}(\theta,\phi)=\epsilon_{\rm M} E_0 R^3\bq~, 
\end{aligned}
\end{equation}
with
\begin{equation}\label{eq:q}
\begin{aligned}
\bq&= \frac{\epsilon_\parallel - \epsilon_{\rm M}}{\epsilon_\parallel+2\epsilon_{\rm M}} \cos(\beta) \bm{e}_z \\
&+ \sum_{\nu=\pm}\frac{\epsilon_\parallel +\nu \epsilon_\bot - \epsilon_{\rm M}}{\epsilon_\parallel+\nu \epsilon_\bot+2\epsilon_{\rm M}} \frac{e^{-i\nu \varphi}}{\sqrt{2}}\sin(\beta)\bm{e}_\nu~, 
\end{aligned}
\end{equation}
which is decomposed in terms of the eigenvectors of the permittivity tensor, see Eq.~\eqref{eq:eps-eig}. Here, it is shown that the dipole moment $\bp$ aligns with the external electric field only when this field is oriented along one of the eigenvectors of the permittivity tensor.
%
%
According to Eq.~\eqref{eq:WT}, inside the sphere, the electric field is expressed as
\begin{equation}\label{eq:Ein}
\begin{aligned}
\bm{E}_{0-{\rm in}}&={E}_0\Big[ \frac{3 \epsilon_{\rm M} }{\epsilon_\parallel+2\epsilon_{\rm M}} \cos(\beta) \bm{e}_z \\
&+ \sum_{\nu=\pm}\frac{3 \epsilon_{\rm M}}{\epsilon_\parallel+\nu \epsilon_\bot+2\epsilon_{\rm M}} \frac{e^{-i\nu \varphi}}{\sqrt{2}}\sin(\beta)\bm{e}_\nu\Big]~, 
\end{aligned}
\end{equation}
which is uniform but generally not aligned with the direction of the field $\bm{E}_0$ or the dipole $\bp$, except if $\bm{E}_0$ is aligned with one of the eigenvectors of the permittivity tensor.

The electric field mentioned above is derived using the permittivity tensor $\epsilon$, whose dependence on $\omega$ is dictated by the microscopic model presented in Sect.~\ref{sec:model}.
In static case $\omega=0$, by using the $\omega$ dependence in $\epsilon_\parallel$ and $\epsilon_\bot$, Eq.~\eqref{eq:WT} gives $W_m=3S_m$ and $T_m=-S_mR^3$. 
Both in the absence of dissipation and for small, yet finite, $\gamma$, one readily obtains $\bq = \bu$, this means that the dipole moment is aligned with the electric field $\bm{E}_0$.
This result is predictable since the unusual polarization directly correlates with the magnetic field, resulting from an oscillating electric field \cite{zyuzin_prb_2012}. Following, we use these findings to examine the effect of an oscillating electric field with a slow dynamics, $0<\omega \ll c/(\sqrt{\epsilon_{\rm M}}R)$, on a WSM nanosphere.

\section{Dipole normal modes of sub-wavelength of a WSM sphere}
\label{sec:quasi}

\begin{figure*}[t]
\centering
\begin{overpic}[width=\columnwidth]{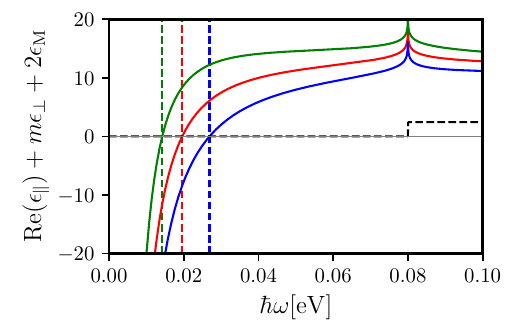}\put(0,60){a)}\end{overpic}\vspace{0em}
\begin{overpic}[width=\columnwidth]{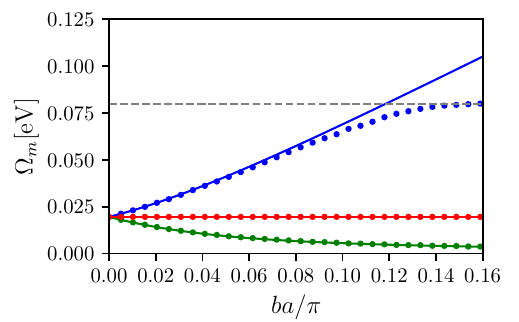}\put(0,60){b)}\end{overpic}\vspace{0em}
\caption{\justifying 
Dipole surface plasmonic resonance conditions. a) $\Re(\epsilon_{\parallel})+m \epsilon_\bot+2 \epsilon_{\rm M}$ are shown as a function of $\hbar \omega$, with $m=-1$ (blue solid line), $m=0$ (red solid line), and $m=1$ (green solid line). Black dashed line depicts $\Im(\epsilon_\parallel)$ as a function of $\hbar \omega$.
The vertical dashed lines represent $\Omega_m$, defined in Eq.~\eqref{eq:dipres}, with $m=-1$ (blue), $m=0$ (red), $m=+1$ (green), and locate the energies which satisfy the 
 resonance condition in Eq.~\eqref{eq:dfrolich}.
Here, we set $b=0.02 \pi/a$.
b) Plasmonic resonance energies as a function of the axion term $b$ in the range $[0,0.16]\pi/a$. Circles represent the numerical solution of Eq.~\eqref{eq:dfrolich} and solid lines are the analytical expressions defined in Eq.~\eqref{eq:dipres}, with $m=-1$ (blue), $m=0$ (red), $m=+1$ (green).
The horizontal gray dashed line represents the threshold $2\epsF$, where $\Im(\epsilon_\parallel)>0$.
In all panels, the calculations have been conducted with parameters $\epsilon_{\rm b}=5$, $\varepsilon_{\rm F}=40$~meV, $v_{\rm D}=c/1000$,  $\Lambda= \hbar v_{\rm D} \pi/a$, $a=3.5$~\AA, and $\epsilon_{\rm M}=1$.
\label{fig:dipole}}
\end{figure*}

Here, we analyze the interaction between an oscillating electromagnetic field and a small WSM spherical particle of radius $R$, such that $R \ll c/(\sqrt{\epsilon_{\rm M}}\omega)$. This implies that the size of the particle is significantly smaller than the wavelength of light in the medium, and one can resort to the quasistatic approximation~\cite{bohren_book}.
The incident oscillating electric field is expressed as a harmonic and uniform $\bm{\mathcal{E}}_{0}=\bm{E}_0 e^{-i \omega t}$, and linearly polarized along the direction $\bm{u}$, see Eq.~\eqref{eq:E0}. 
The associated magnetic field is expressed as $\bm{\mathcal{B}}_{0}=\bm{B}_0 e^{-i \omega t}$, where
\begin{equation}\label{eq:B0}
\bm{B}_0 =  \frac{i \omega \epsilon_{\rm M}  }{2 c}   (\br \times \bm{E}_{0})~,
\end{equation}
which describes a transverse magnetic (TM) field, and
such that $\nabla \times \bm{B}_{0}=-i (\omega/ c)\bm{D}_0$, with $\bm{D}_0=\epsilon_{\rm M} \bm{E}_0$.
Within the quasistatic approximation, we begin with the electrostatic issue addressed in Section~\ref{sec:electrostatics}. The incident electric field generates an oscillating electric dipole $\bp e^{-i\omega t}$, with $\bp$ as defined in Eq.~\eqref{eq:p}.
In the near zone $r \ll c/(\sqrt{\epsilon_{\rm M}}\omega)$, an oscillating electric dipole generates the following electric and magnetic fields~\cite{Jackson:100964} 
\begin{subequations}
\begin{align}
\bm{\mathcal{E}}_{0-\rm sc-out}&\approx \bm{E}_{\rm dip} e^{-i \omega t}~,\\
\bm{\mathcal{B}}_{0-\rm sc-out}&\approx \frac{i \omega }{c} \frac{\bm{e}_r \times \bm{p} e^{-i \omega t}}{r^2}~,
\end{align}
\end{subequations}
which are linked by the relation $\nabla \times \bm{\mathcal{B}}_{0-\rm sc-out}=c^{-1}\frac{\partial \bm{\mathcal{D}}_{0-\rm sc-out}}{\partial t}$, where $\bm{\mathcal{D}}_{0-\rm sc-out}=\epsilon_{\rm M}\bm{\mathcal{E}}_{0-\rm sc-out}$. 
%
Similarly, inside the sphere, the electric field has a homogeneous spatial form
\begin{equation}
\bm{\mathcal{E}}_{0-\rm sc-in}\approx \bm{E}_{0-{\rm in}} e^{-i \omega t},
\end{equation}
where $\bm{E}_{0-{\rm in}}$ is defined in Eq.~\eqref{eq:Ein}, and the magnetic field is expressed as
\begin{equation}
\bm{\mathcal{B}}_{0-\rm sc-in}\approx\frac{i \omega }{2 c} (\br \times \bm{D}_{0-{\rm in}} e^{-i \omega t} )
\end{equation}
where $\bm{D}_{0-{\rm in}}=\epsilon \cdot\bm{E}_{0-{\rm in}}$, 
and it is valid the relation $\nabla \times \bm{\mathcal{B}}_{0-\rm sc-in}=\frac{1}{c}\frac{\partial \bm{\mathcal{D}}_{0-\rm sc-in}}{\partial t}$,
where $\bm{\mathcal{D}}_{0-\rm sc-in}\approx \bm{D}_{0-\rm in} e^{-i \omega t}$.
It is observed that, within the quasistatic approximation, the continuity of the magnetic field across the surface of the sphere is preserved.

Now, we discuss the absorption of the incoming light by the spherical particle. For this purpose, we analyze the Poynting vector associated with the time-harmonic field on a closed surface that surrounds the sphere~\cite{bohren_book}. 
The Poynting vector averaged in time for the time-harmonic field can be decomposed as
\begin{equation}\label{eq:SSS}
{\bm S}_{0-\rm out}={\bm S}_{0-\rm inc}+{\bm S}_{0-\rm sc}+{\bm S}_{0-\rm ext}~,
\end{equation}
where the contribution due to the incident wave is
\begin{equation}
{\bm S}_{0-\rm inc}= \frac{c}{8\pi} \Re[\bm{\mathcal{E}}_{0} \times \bm{\mathcal{B}}_{0}^\ast ]~,
\end{equation}
the term due to the scattered field is 
\begin{equation}
{\bm S}_{0-\rm sc}=\frac{c}{8\pi}  \Re[\bm{\mathcal{E}}_{\rm 0-sc-out} \times  \bm{\mathcal{B}}_{\rm 0-sc-out}^\ast ]~,
\end{equation}
and the last contribution is
\begin{equation}
\bm{S}_{0-\rm ext} = \frac{c}{8\pi}  \Re[
\bm{\mathcal{E}}_{0}     
\times  \bm{B}_{0-\rm sc-out}^\ast +\bm{\mathcal{E}}_{0-\rm sc-out} \times  \bm{\mathcal{B}}_{0}^\ast     ]~,
\end{equation}
which comes from the interaction between the incident and scattered waves.
The total rate at which the electromagnetic energy traverses a spherical surface of radius $R$ centered at the particle is
%
\begin{equation}
W_0=\int d^3 \bm{r} \delta(r-R) \bm{e}_r\cdot {\bm S}_{0-\rm out}~.
\end{equation}
%
Using the decomposition in Eq.~\eqref{eq:SSS}, we split the total electromagnetic power into three terms $W_0=W_{0-\rm inc}+W_{0-\rm sc}-W_{0-\rm ext}$.
Within the quasistatic approximation, we evaluate each contribution by examining the electromagnetic power and focusing on the lowest order term in an expansion parameterized by $\sqrt{\epsilon_{\rm M}}(\omega R/c)$. 
%
It is known that for any particle~\cite{bohren_book}, the contribution $W_{\rm inc}$ is zero if the medium is not absorbing.
Indeed, the incident electromagnetic power is expressed as
\begin{equation}
\begin{aligned}
W_{0-\rm inc}&=-\int d^3 \bm{r} \delta(r-R) \bm{e}_r\cdot {\bm S}_{0-\rm inc}\\
&\approx\frac{-\omega \Im(\epsilon_{\mathrm M}) R^3 |E_0|^2}{6}~,
\end{aligned}
\end{equation}
which is exactly zero because $\Im(\epsilon_{\rm M})=0$.
Similarly, we determine the energy absorption rate, 
resulting in
\begin{equation}
\begin{aligned}
W_{0-\rm sc}&=\int d^3 \bm{r} \delta(r-R) \bm{e}_r\cdot {\bm S}_{0-\rm sc}\\
&\approx
\frac{\omega \Im(\epsilon_{\rm M}) R^3 |E_0|^2 |\bm{q}|^2 }{3}~,
\end{aligned}
\end{equation}
which is negligible if the medium is not absorbing. 
Following a textbook approach~\cite{bohren_book}, to estimate the lowest nonzero correction in $\sqrt{\epsilon_{\rm M}}\omega R/c$ of the energy absorption rate, we consider the electromagnetic field generated by an oscillating dipole $\bp=\epsilon_{\rm M} E_0 R^3 \bq$ in the far zone $\sqrt{\epsilon_{\rm M}}\omega r/c\gg 1$, that is, 
\begin{subequations}
\begin{align}
{\bm B}_{\rm far-dip}&=\epsilon_{\rm M}\frac{\omega^2}{c^2} \bm{e}_r \times \bp \frac{e^{i\sqrt{\epsilon_{\rm M}}\frac{\omega}{c}r}}{r}~,\\
\bm{E}_{\rm far-dip}&=\frac{1}{\sqrt{\epsilon_{\rm M}}}
 {\bm B}_{\rm far-dip} \times \bm{e}_r~,
\end{align}
\end{subequations}
where the associated Poynting vector averaged in time for the time harmonic is ${\bm S}_{\rm far-dip}=\frac{c}{8 \pi}\Re[{\bm E}_{\rm far-dip} \times {\bm B}_{\rm far-dip }^\ast]$. 
The associated energy rate is given by
\begin{equation}
\begin{aligned}
W_{\rm far-dip} &=  \lim_{r_\infty \to \infty} \int d^3 \br \delta(r-r_\infty) \bu_r\cdot {\bm S}_{\rm far-dip} \\
&= \frac{ \epsilon_{\rm M}^{3/2} \omega^4}{3 c^3} |E_0|^2  R^6 |\bm{q}|^2 ,
\end{aligned}
\end{equation}
which indicates an estimate for the rate at which electromagnetic energy is scattered in the regime where $\sqrt{\epsilon_{\rm M}}\omega R/c$ is small, and follows the scaling law $(\sqrt{\epsilon_{\rm M}} \omega R/c)^3$.
%
Finally, within the quasistatic approximation,  the extinction electromagnetic power is expressed as 
\begin{equation}\label{eq:Wext}
\begin{aligned}
W_{0-\rm ext}&=-\int d^3 \bm{r} \delta(r-R) \bm{e}_r\cdot {\bm S}_{0-\rm ext}\\
&\approx\frac{\omega \Im(\epsilon_{\rm M}^\ast q_u) R^3 |E_0|^2}{2}~,    
\end{aligned} 
\end{equation}
where $q_u=\bm{q}\cdot \bm{u}$ is the projection of the dipole moment along the direction of the incident electric field in units of $\epsilon_{\rm M} R^3 E_0$.
%
%
For a sub-wavelength sphere inside a non-absorbing medium, the extinction is primarily influenced by absorption, such that
$W_0\approx W_{0-\rm ext}$, for which the absorption cross section is given by

\begin{equation}
C_{\rm abs}=\frac{W_0}{I_0}\approx 4 \pi \frac{\omega \sqrt{\epsilon_{\rm M}}}{c} R^3 \Im(q_u)~,
\end{equation}
where $I_0=\frac{c\sqrt{\epsilon_{\rm M}}}{8 \pi} |E_0|^2$~\cite{akimov_book,bohren_book}. 
In contrast to a small sphere composed of an isotropic metallic material, here the vector $\bq$ is not aligned with the incident electric field and this misalignment introduces a non-trivial relation between the absorption cross section and the polarization of the incident electromagnetic field 
\begin{equation}
\begin{aligned}
\Im(q_u)
&
=\Im\Big[\frac{\epsilon_\parallel-\epsilon_{\rm M}}{2\epsilon_\parallel+\epsilon_{\rm M}}\Big] \cos^2(\beta) \\
&
+
\Im\Big[\sum_{\nu=\pm}\frac{\epsilon_\parallel +\nu \epsilon_\bot - \epsilon_{\rm M}}{\epsilon_\parallel+\nu \epsilon_\bot+2\epsilon_{\rm M}}\Big] \frac{\sin^2(\beta)}{2}~.
\end{aligned}
\end{equation}
For small values of $|\Im(\epsilon_\parallel)| \ll 1$, the following condition 
\begin{equation}\label{eq:dfrolich}
\Re(\epsilon_\parallel) + m\epsilon_\bot + 2 \epsilon_{\rm M} = 0~,
\end{equation}
with $m\in\{0,\pm1\}$,
determines the resonance excitation energies for dipole symmetric LSP. 
When $\Im(\epsilon_\parallel)=0$, Eq.~\eqref{eq:dfrolich} indicates that the excitation has an infinite lifetime. 
This resonance condition allows an evanescent external field to generate a self-sustaining electric field, and identify specific energies where enhanced electromagnetic absorption takes place. 
This scenario is similar to the dipole Fr\"ohlich condition obtained in standard metals~\cite{hayashi_jpd_2012}. 
However, due to the axion term $\epsilon_\bot$, here, there is an increase in the number of excitation energies. 
The mechanism is similar to that of bulk plasmons in WSMs~\cite{pellegrino_prb_2015}, where the axion term breaks the degeneracy of the three gapped collective modes in the long-wavelength limit.
In Eq.~\eqref{eq:dfrolich}, each index $m$ refers to an eigenvector of the permittivity tensor. 
Thus, when the resonance condition holds for a particular $m$, the self-sustained electric field is generated by an oscillating dipole moment aligned with the involved eigenvector of the permittivity tensor. Specifically, for $m=0$ ($m=\pm 1$), it aligns along $\bm{e}_z$ ($\bm{e}_{\pm}$). 
Recently~\cite{hu_prmat_2023,naeimi_prmat_2025}, the result in Eq.~\eqref{eq:dfrolich} has been explored to analyze the mechanism of near-field heat-transfer rectification between two WSM nanoparticles and a planar WSM substrate via the coupling to nonreciprocal surface modes.
%
In particular, the eigenmodes labeled by $m=-1$ and $m=+1$ correspond, respectively, to a dipole and a uniform electric field inside the sphere, with opposite circular polarizations. These fields lie in the plane orthogonal to the $\bm{b}$ vector and have two distinct proper frequencies. 
This behavior resembles the emergence of nonreciprocal surface plasmons in Weyl semimetals~\cite{hofmann_prb_2016}, where the electric fields lie in the orthogonal plane to the vector $\bm{b}$.
The circular polarization of the dipole modes with $m = \pm 1$ distinguishes the case of a WSM nanosphere from that of a nanosphere composed of a generic anisotropic material, which does not present a Hall-like response. In the latter case, the permittivity tensor takes the diagonal form 
$\epsilon=\epsilon_{x'x'} \bm{e}_{x'}\otimes\bm{e}_{x'}+\epsilon_{y'y'} \bm{e}_{y'}\otimes\bm{e}_{y'}+\epsilon_{z'z'} \bm{e}_{z'}\otimes\bm{e}_{z'}$
and $\{ \bm{e}_{x'}, \bm{e}_{y'}, \bm{e}_{z'} \}$ constitute an orthonormal basis. Here, the dipole eigenmodes are dictated by the condition $\Re[\epsilon_{jj}]+2 \epsilon_{\rm M}=0$,
where $j = x', y', z'$, and each mode is linearly polarized along the corresponding principal axis. 

In Fig.~\ref{fig:dipole}~a), the conditions for dipole surface plasmonic resonance as described in Eq.~\eqref{eq:dfrolich} are illustrated for a set of microscopic parameters reported in the caption.
Contrary to a conventional metal described by the Drude model~\cite{maier_book}, we observe that the resonant condition is satisfied by three distinct energies rather than a single excitation energy.
In this analysis, the impact of the interband contribution to optical conductivity can be effectively described through a renormalization of the dielectric constant $\epsilon_{\rm b}$. Essentially, modeling the interband effect on the permittivity tensor as a modification of $\epsilon_{\rm b}\to\epsilon_{\rm b}^\star(w_0)$, where
\begin{equation}\label{eq:eps*}
\epsilon_{\rm b}^\star(w_0) = \epsilon_{\rm b} + \frac{\alpha c}{3 \pi v_{\rm D}} \log\Big| \frac{4 \Lambda^2}{4\epsF^2 - w_0^2}\Big|~,
\end{equation}
and $w_0=\hbar \omega_{\rm pl}/\sqrt{\epsilon_{\rm b}+2\epsilon_{\rm M}}$ represents the dipole localized plasmonic energy without the axion term and interband optical conductivity~\cite{pellegrino_prb_2015}. So, by replacing $\epsilon_\parallel \to  \epsilon_{\rm b}^\star(w_0)-\frac{\omega_{\rm pl}^2}{\omega(\omega + i\gamma)}$  in Eq.~\eqref{eq:dfrolich}, and taking the ultraclean limit $\gamma=0^+$, we derive straightforward analytical expressions for the plasmonic resonance energies
\begin{equation}\label{eq:dipres}
\begin{aligned}
\Omega_{m}&=\sqrt{\frac{\hbar^2\omega_{\rm pl}^2}{\epsilon_{\rm b}^\star(w_0)+2 \epsilon_{\rm M}}+\bigg[\frac{m \alpha \hbar c b}{\pi (\epsilon_{\rm b}^\star(w_0)+2 \epsilon_{\rm M})}\bigg]^2}\\
&-\frac{m \alpha \hbar c b}{\pi (\epsilon_{\rm b}^\star(w_0)+2 \epsilon_{\rm M})}~.
\end{aligned}
\end{equation}
For the fixed value $b=0.02\pi/a$, these energies are illustrated in Fig.~\ref{fig:dipole}~a) by vertical dashed lines, and precisely indicate the conditions for dipole surface resonance, which are located well below $2 \varepsilon_{\rm F}$, the threshold for interband transitions, such that $\Im(\epsilon_\parallel)=0$ which ensures a infinite excitation lifetime in ultraclean limit.
Fig.~\ref{fig:dipole}~b) shows the plasmonic resonance energies, calculated numerically by solving Eq.~\eqref{eq:dfrolich} (depicted by circles), as a function of $b$. In comparison, we present the analytical expressions detailed in Eq.~$\eqref{eq:dipres}$ (shown as solid lines), which exhibit a good agreement with numerical results. We note a visible discrepancy between the analytical and numerical outcomes only when the plasmonic resonant energies approach the threshold energy $2 \epsF$, indicated by the horizontal dashed gray line.

%
\section{A WSM sphere in an electric field with a linear spatial dependence}
\label{sec:electrostatics-q}

Following the approach of Sect.~\ref{sec:electrostatics}, we consider a WSM sphere of radius $R$, in the presence of a static electric field with a linear spatial dependence.
The static electric field can be decomposed in terms of vector spherical harmonics~\cite{barrera_epj_1985} as follows:
\begin{equation}\label{eq:EQstatic}
\bm{E}_{\rm Q}=\sum_{m=-2}^2\frac{E_{{\rm Q},m} r}{R} [2  \bm{Y}_{2 m}(\theta,\phi)
+ \bm{\Psi}_{2 m}(\theta,\phi)]~,
\end{equation}
which is solenoidal, $\nabla \cdot \bm{E}_{\rm Q}=0$, and irrotational, $\nabla \times\bm{E}_{\rm Q}=0$, for any set of coefficients $\{E_{{\rm Q},m} \}$. 
Here, the electric displacement
field is ${\bm D}_{\rm Q}=\epsilon_{\rm M} \bm{E}_{\rm Q}$.
The aforementioned electrostatic field represents a simple example of non-trivial spatial dependence.
As an example, we consider a planar surface charge distribution situated at $z=L_z$, given by $\rho_{\rm 2D}(x)=\rho_0  x$,  a second planar conductor at $z=0$ forced at ground potential, and they are embedded in a medium with a simple dielectric constant $\epsilon_{\rm M}$.
This setup results in an electrostatic potential described by $\psi=-\frac{\rho_0}{2 \epsilon_{\rm M}L_z} {\rm sgn}(z-L_z)x z$, where $
{\rm sgn}(z)$ is the sign function.
Within the region $0<z<L_z$, the electric field is given by $-\nabla \psi =-\frac{\rho_0 }{2 \epsilon_{\rm M}L_z} (z \bm{e}_x + x \bm{e}_z)$. This is a particular case of the electric field in Eq.~\eqref{eq:EQstatic}, and it involves only the azimuthal labels $m=\pm1$.
%

Starting from the general electrostatic field with a linear spatial dependence in Eq.~\eqref{eq:EQstatic}, we employ an Ansatz for the electric field inside the sphere 
\begin{equation}\label{eq:EQin_0}
\bm{E}_{\rm in}=\bm{E}'_{\rm in}+\bm{E}''_{\rm in}+\bm{E}'''_{\rm in}~,    
\end{equation}
with
\begin{subequations}\label{eq:EQin}
\begin{align}
\bm{E}'_{\rm in}&=r\sum_{m=-1}^1 X'_m \bm{\Phi}_{1m}(\theta,\phi)~,\label{eq:EQinp}\\
\bm{E}''_{\rm in}&=r\sum_{m=-2}^2 X''_m [2\bm{Y}_{2m}(\theta,\phi)+ \bm{\Psi}_{2m}(\theta,\phi)]~,
\label{eq:EQinpp}\\
\bm{E}'''_{\rm in}&=r X'''  {\bm Y}_{00}(\theta,\phi)~,\label{eq:EQinppp}
\end{align}
\end{subequations}
where we highlight that $\bm{E}''_{\rm in}$ takes the form presented in Eq.~\eqref{eq:EQstatic}.
Applying the permittivity tensor to the electric field in Eq.~\eqref{eq:EQin_0}, the electric displacement vector is expressed as 
\begin{equation}
\bm{D}_{\rm in}=\epsilon \cdot \bm{E}_{\rm in}=\bm{D}'_{\rm in}+\bm{D}''_{\rm in}+\bm{D}'''_{\rm in}~,    
\end{equation}
with
\begin{subequations}\label{eq:DQin}
\begin{align}
\bm{D}'_{\rm in}&=r\sum_{m=-1}^1 X'_{D m} \bm{\Phi}_{1m}(\theta,\phi)~,\\
\bm{D}''_{\rm in}&=r\sum_{m=-2}^2 X''_{D m} [2\bm{Y}_{2m}(\theta,\phi)+ \bm{\Psi}_{2m}(\theta,\phi)]~,
\\
\bm{D}'''_{\rm in}&=r X_D'''  {\bm Y}_{00}(\theta,\phi)~,
\end{align}
\end{subequations}
where the same set of vector spherical harmonics of $\bm{E}_{\rm in}$ are involved, and one has
\begin{subequations}
\begin{align}
\begin{pmatrix}
X_{D 0}'\\
X_{D 0}''\\
X'''_D
\end{pmatrix}
&=
\begin{pmatrix}
\epsilon_\parallel & i\epsilon_\bot \sqrt{\frac{5}{3}}& \frac{-i }{\sqrt{3}} \epsilon_\bot\\
-\frac{i}{\sqrt{15}}  \epsilon_\bot&\epsilon_\parallel&0\\
i\frac{2}{\sqrt{3}}  \epsilon_\bot&0 & \epsilon_\parallel
\end{pmatrix}
\begin{pmatrix}
X_{ 0}'\\
X_{ 0}''\\
X'''
\end{pmatrix}~,
\\
\begin{pmatrix}
X_{D \pm 1}'\\
X_{D \pm 1}''
\end{pmatrix}
&=
\begin{pmatrix}
\epsilon_\parallel \pm \frac{1}{2} \epsilon_\bot & i \frac{\sqrt{5}}{2} \epsilon_\bot \\
- \frac{i}{2\sqrt{5}} \epsilon_\bot & \epsilon_\parallel \pm \frac{1}{2}\epsilon_\bot
\end{pmatrix}
\begin{pmatrix}
X_{ \pm 1}'\\
X_{ \pm 1}''
\end{pmatrix}~,\\
X''_{D \pm2}&=(\epsilon_\parallel\pm \epsilon_\bot) X''_{\pm 2}~.
\end{align}
\end{subequations}
%
%
For Gauss's law to hold for the electric displacement vector inside the sphere, as highlighted in Eq.~\eqref{eq:nablaD}, it is necessary that $X'''_{D }=0$. This requirement is equivalent to the following condition
\begin{equation}\label{eq:Xppp}
\epsilon_\parallel X''' = \frac{-i2\epsilon_\bot}{\sqrt{3}} X_0'~.
\end{equation}
Similarly, we write the electric field outside the sphere as $\bm{E}_{\rm Q}+\bm{E}_{\rm out}'+\bm{E}_{\rm out}''$,
%
%
where the electric field $\bm{E}_{\rm Q}$ is augmented by the following contributions
\begin{subequations}\label{eq:EQout}
\begin{align}
\bm{E}_{\rm out}'&=\sum_{m=-1}^1 \frac{Z'_m}{r^2}  \bm{\Phi}_{1m}(\theta,\phi)~, \label{eq:Epout}\\
\bm{E}_{\rm out}''&=  \sum_{m=-2}^{2} \frac{Z''_m R^2}{r^4}[-3 \bm{Y}_{2m}(\theta,\phi)+\bm{\Psi}_{2m}(\theta,\phi)]~,
\label{eq:qstaticfield}
\end{align}
\end{subequations}
for which two electric displacement vectors are defined: $\bm{D}'_{\rm out}=\epsilon_{\rm M}\bm{E}'_{\rm out}$ and $\bm{D}''_{\rm out}=\epsilon_{\rm M} \bm{E}'_{\rm out}$, both of which comply with Gauss's law.
%
To ensure the continuity of the electric field vector components tangential to the surface of the sphere, as outlined in Eq.~\eqref{eq:bcE}, we obtain
\begin{subequations}\label{eq:contEQ}
\begin{align}
X_m''&=\frac{Z_m''}{R^3}+\frac{E_{{\rm Q},m}}{R}~,\\
X_m' &=\frac{Z_m'}{R^3}~. \label{eq:contEQb}
\end{align}
\end{subequations}
By enforcing the continuity condition of the electric displacement vector component normal to the surface of the sphere, as indicated in Eq.~\eqref{eq:bcD}, we arrive at the expressions below
\begin{equation}\label{eq:contDQ}
X_{Dm}''=\epsilon_{\rm M}\Big( \frac{E_{Q,m}}{R}-\frac{3}{2}\frac{Z_m''}{R^2}\Big)~.
\end{equation}
To satisfy Faraday's law,  in the static limit $\nabla \times \bm{E}=0$, we have to impose that the electric field is irrotational both inside and outside the sphere, which means $X'''=0$ and $X_m'=Z'_m=0$ for any $m\in\{0,\pm1\}$, and the only nonzero coefficients are
\begin{subequations}\label{eq:EQoutstatic}
\begin{align}
X''_m&=\frac{5\epsilon_{\rm M}}{2\epsilon_\parallel+m\epsilon_\bot+3\epsilon_{\rm M}} \frac{ E_{{\rm Q},m}}{R}~,\\
Z''_m&=\frac{2\epsilon_{\rm M}-(2\epsilon_\parallel+m\epsilon_\bot)}{2\epsilon_\parallel+m\epsilon_\bot+3\epsilon_{\rm M}} R^2 E_{{\rm Q},m}~.
\end{align}
\end{subequations}
%
%
%
Outside the sphere, the electric field is $\bm{E}_{\rm Q}+\bm{E}_{\rm out}''$. The electric field generated by the response of the WSM $\bm{E}_{\rm out}''$ has the form of an electrostatic field $-\nabla V_{\rm qp}(\br)$, where the electrostatic potential is given by
\begin{equation}\label{eq:VQ}
V_{\rm qp}(\br) = \frac{4 \pi}{5} \sum_{m=-2}^2 \frac{Q_{2m}}{r^3} Y_{2m}(\theta,\phi)~,
\end{equation}
arising from quadrupole moments~\cite{Jackson:100964}. Here, the quadrupole moments are 

\begin{equation}\label{eq:Q2m}
Q_{2m}=-\frac{5R^2}{4 \pi}  Z_m''~.
\end{equation}
%
In conclusion, we explicitly write the terms that define the electric displacement vectors inside the sphere as
\begin{subequations}\label{eq:XD_TM}
\begin{align}
X'_{D 0}&=i \epsilon_\bot \sqrt{\frac{5}{3}}\frac{5\epsilon_{\rm M}}{2\epsilon_{\parallel}+3\epsilon_{\rm M}} \frac{ E_{{\rm Q},0}}{R}~\\  
X'_{D \pm 1}&=i \epsilon_\bot \frac{\sqrt{5}}{2}\frac{5\epsilon_{\rm M}}{2\epsilon_{\parallel}+m \epsilon_\bot+3\epsilon_{\rm M}} \frac{ E_{{\rm Q},\pm 1}}{R}~,\\
X''_{D m}&=\frac{5 \epsilon_{\rm M}(\epsilon_{\parallel}+m \epsilon_\bot/2)}{
2 \epsilon_{\parallel}+m \epsilon_\bot+3 \epsilon_{\rm M}} \frac{E_{{\rm Q}m}}{R}~,
\end{align}
\end{subequations}
with $m\in \{0,\pm1,\pm2\}$.

The aforementioned electric field is determined using the permittivity tensor $\epsilon$, whose dependence on $\omega$ follows the microscopic model described in Sect.~\ref{sec:model}.
In the static case $\omega=0$, employing the $\omega$ dependence in $\epsilon_\parallel$ and $\epsilon_\bot$, the coefficients in Eq.~\eqref{eq:EQoutstatic} are $Z_m''=-R^2 E_{{\rm Q},m}$ and $X_m''=0$, indicating that the electric field is zero inside the sphere.

\section{Quadrupole normal modes of a sub-wavelength WSM sphere}
\label{sec:quasi-q}

In analogy to Sect.~\ref{sec:quasi}, we introduce an oscillating electromagnetic field, described in the quasistatic approximation. We analyze the interaction between a small spherical WSM particle and an oscillating electric field expressed as $\bm{\mathcal{E}}_{\rm Q}={\bm E}_{\rm Q} e^{-i \omega t}$, where  ${\bm E}_{\rm Q}$ is defined in Eq.~\eqref{eq:EQstatic} and has a linear spatial dependence.
The associated magnetic field is $\bm{\mathcal{B}}_{\rm Q}={\bm B}_{\rm Q} e^{-i \omega t}$, where
\begin{equation}\label{eq:BQ}
{\bm B}_{\rm Q}=i \frac{\omega}{c} \frac{\epsilon_{\rm M}}{R}\frac{r^2}{3} \sum_{m=-2}^2E_{{\rm Q},m} \bm{\Phi}_{2 m}(\theta,\phi),
\end{equation}
so that $\nabla \times \bm{B}_{\rm Q}=-i \omega/c \bm{D}_{\rm Q}$ and $\bm{D}_{\rm Q}=\epsilon_{\rm M}\bm{E}_{{\rm Q}}$, and it describes a TM field.
Inside the sphere, the scattered electric field generated has the form $\bm{\mathcal{E}}_{\rm Q-sc-in}\approx  \bm{E}_{\rm in}''e^{-i \omega t}$, defined in Eq.~\eqref{eq:EQinpp}. The associated magnetic field generated inside the sphere can be approximated as $\bm{\mathcal{B}}_{\rm Q- sc-in}\approx ({\bm B}_{\rm in}' +{\bm B}_{\rm in}'') e^{- i \omega t}$, where
\begin{subequations}\label{eq:Bin}
\begin{align}
\bm{B}_{\rm in}'&=-i\frac{\omega}{ c} \sum_{m=-1}^1    X_{D m}'  \Big[\Big(\frac{r^2}{5}-\frac{R^2}{3}\Big)\bm{Y}_{1 m}(\theta,\phi)\nonumber\\
&~~~+\Big(\frac{2r^2}{5}-\frac{R^2}{3}\Big)\bm{\Psi}_{1 m}(\theta,\phi)\Big]~,\\
\bm{B}_{\rm in}''&=i\frac{\omega}{ c} \frac{r^2}{3}\sum_{m=-2}^2    X_{D m}''  \bm{\Phi}_{2m}(\theta,\phi)~. \label{eq:Binpp}
\end{align}
\end{subequations}
Here, $\bm{B}_{\rm in}'$ is a dipole symmetric  field, such that $\nabla \times {\bm B}_{\rm in}'=-i (\omega/c) \bm{D}_{\rm in}'$, and $\bm{B}_{\rm in}''$ is a quadrupole symmetric field, such that $\nabla \times {\bm B}_{\rm in}''=-i (\omega/c) \bm{D}_{\rm in}''$.
Outside the sphere, in the near zone $r \ll c/(\sqrt{\epsilon_{\rm M}}\omega)$, the electric field generated as the response of the sphere has the form $\bm{\mathcal{E}}_{\rm Q-sc-out}\approx\bm{E}_{\rm out}''e^{-i \omega t}$, defined in Eq.~\eqref{eq:EQout}. The associated magnetic field generated outside the sphere can be approximated as
$
\bm{\mathcal{B}}_{\rm Q-sc-out}\approx (  {\bm B}_{\rm out}'+{\bm B}_{\rm out}'' )e^{- i \omega t},
$
where
\begin{subequations}\label{eq:Bout}
\begin{align}
\bm{B}_{\rm out}'&=-i\frac{\omega}{c} \sum_{m=-1}^1\frac{X'_{D m }R^4}{15r^2}
[-2\bm{Y}_{1 m}(\theta,\phi)+\bm{\Psi}_{1 m}(\theta,\phi)]~,\label{eq:Bpout}\\
\bm{B}_{\rm out}''&=-i\frac{\epsilon_{\rm M} \omega R^2}{c} \sum_{m=-2}^2 \frac{Z''_m }{2r^3}   \bm{\Phi}_{2m}(\theta,\phi)~. \label{eq:Boutpp}
\end{align}
\end{subequations}
Here, $\bm{B}_{\rm out}'$ is a dipole symmetric field, such that  $\nabla \times {\bm B}_{\rm out}'=-i (\omega/c) \bm{D}_{\rm out}'$, $\bm{B}_{\rm out}'' $ is a quadrupole symmetric TM field, such that $\nabla \times {\bm B}_{\rm out}''=-i (\omega/c) \bm{D}_{\rm out}''$.
Moreover, the continuity of the magnetic field across the surface of the sphere is verified.
To analyze light absorption by a spherical particle, we look into the Poynting vector for the time-harmonic field over a surface enclosing the sphere, as outlined in Eq.~\eqref{eq:SSS}~\cite{bohren_book}. The electromagnetic extinction power $W_{\rm Q-ext}=-\int d^3 \bm{r} \delta(r-R) \bm{e}_r\cdot {\bm S}_{\rm Q-ext}$ is connected with the Poynting vector 
\begin{equation}
{\bm S}_{\rm Q-ext}\approx \frac{c}{8\pi} 
  \Re[
\bm{\mathcal{E}}_{\rm Q}   
\times  \bm{\mathcal{B}}_{\rm Q-sc-out}^\ast +\bm{\mathcal{E}}_{\rm Q-sc-out} \times  \bm{\mathcal{B}}_{\rm Q} ^\ast     ]~,
\end{equation}
such that, in the case $\sqrt{\epsilon_{\rm M}}\omega R/c\ll 1$, one has
\begin{equation}\label{eq:WQext}
\begin{aligned}
 W_{\rm Q-ext}&\approx \frac{5}{4\pi} \omega R^3 \epsilon_{\rm M} \sum_{m=-2}^2\Big[ |E_{{\rm Q},m}|^2 \\
 &\times \Im\bigg( \frac{\epsilon_{\parallel}+\epsilon_\bot/2-\epsilon_{\rm M}}{2\epsilon_{\parallel}+\epsilon_\bot+3\epsilon_{\rm M}} \Big)\bigg]~.   
\end{aligned}
\end{equation}
So, as in Sect.~\ref{sec:quasi}, when dealing with a sub-wavelength sphere within a nonabsorbing medium, the extinction is predominantly controlled by the absorption. Under the quasistatic approximation, the terms of the Poynting vectors of the incoming wave and the scattered fields 
\begin{subequations}
\begin{align}
{\bm S}_{\rm Q-inc}&\approx \frac{c}{8\pi} 
  \Re[
   \bm{\mathcal{E}}_{\rm Q}  \times  \bm{\mathcal{B}}_{\rm Q} ^\ast   ]~,       \\
{\bm S}_{\rm Q-sc}&\approx \frac{c}{8\pi} 
  \Re[
\bm{\mathcal{E}}_{\rm Q-sc-out}  \times  \bm{\mathcal{B}}_{\rm Q-sc-out}^\ast    ]~,
\end{align}
\end{subequations}
produce two vanishing electromagnetic power components, i.e., $W_{\rm Q-inc}=-\int d^3 \br \delta(r-E) \bm{e}_r \cdot {\bm S}_{\rm Q-inc}= 0$ and $W_{\rm Q-sc}=\int d^3 \br \delta(r-E) \bm{e}_r \cdot {\bm S}_{\rm Q-sc}\approx0$, provided $\Im(\epsilon_{\rm M})=0$. 
Thus, for $\sqrt{\epsilon_{\rm M}}\omega R/c\ll 1$, 
the absorption power $W_{\rm Q}=W_{\rm Q-inc}+W_{\rm Q-sc}-W_{\rm Q-ext}\approx W_{\rm Q-ext}$, and  the absorption cross section can be approximated as
\begin{equation}
\begin{aligned}    
C_{\rm Q-abs}=\frac{W_{\rm Q}}{I_{\rm Q}}
&\approx 4 \pi \frac{\omega\sqrt{\epsilon_{\rm M}}}{c} R^3\\
&\times 
\frac{ \sum_{m=-2}^2 |E_{{\rm Q},m}|^2 \Im\Big( \frac{\epsilon_\parallel+m\epsilon_\bot/2-\epsilon_{\rm M}}{2\epsilon_\parallel+m\epsilon_\bot+3\epsilon_{\rm M}} \Big)}{\sum_{m'=-2}^2|E_{{\rm Q},m}|^2}~,
\end{aligned}
\end{equation}
where $I_{\rm Q}= \frac{c \sqrt{\epsilon_{\rm M}}}{8 \pi} \frac{1}{4 \pi}\int d^3 \br \delta(r-R) |{\bm E}_{\rm Q}|^2$.

\begin{figure*}[t]
\centering
\begin{overpic}[width=\columnwidth]{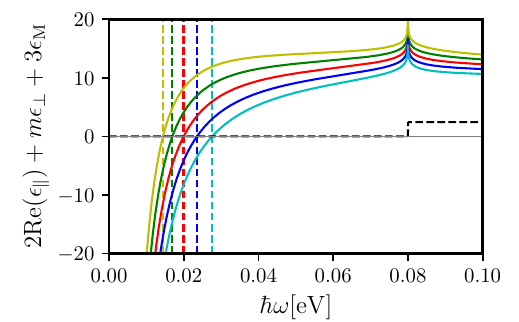}\put(0,60){a)}\end{overpic}\vspace{0em}
\begin{overpic}[width=\columnwidth]{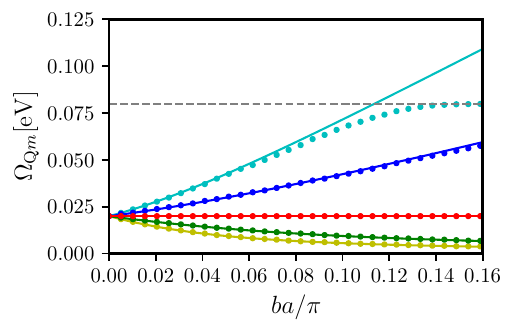}\put(0,60){b)}\end{overpic}\vspace{0em}
\caption{\justifying 
Quadrupole surface plasmonic resonance conditions.
a) $2\Re(\epsilon_{\parallel})+m \epsilon_\bot+3\epsilon_{\rm M}$ are shown as a function of $\hbar \omega$, with $m=-2$ (cyan solid line), $m=-1$ (blue solid line), $m=0$ (red solid line), $m=1$ (green solid line),  and $m=2$ (yellow solid line). Black dashed line is $\Im(\epsilon_{m})=\Im(\epsilon_\parallel)$ as a function of $\hbar \omega$. 
The vertical dashed line represents $ \Omega_{{\rm Q}m}$, defined in Eq.~\eqref{eq:qres}, with $m=-2$ (cyan), $m=-1$ (blue), $m=0$ (red), $m=+1$ (green), $m=+2$ (yellow), and locate the energies which satisfy the 
resonance condition in Eq.~\eqref{eq:qfrolich}.
Here, we set $b=0.02 \pi/a$.
b) Plasmonic resonance energies as a function of the axion term $b$ in the range $[0,0.16]\pi/a$. Circles represent the numerical solution of Eq.~\eqref{eq:qfrolich} and solid lines are the analytical expressions defined in Eq.~\eqref{eq:qres},  with $m=-2$ (cyan), $m=-1$ (blue), $m=0$ (red), $m=+1$ (green), $m=+2$ (yellow).
The horizontal gray dashed line represents the threshold $2\epsF$, where  $\Im(\epsilon_\parallel)>0$.
In all panels, the calculations have been conducted with parameters $\epsilon_{\rm b}=5$, $\varepsilon_{\rm F}=40$~meV, $v_{\rm D}=c/1000$, $\Lambda= \hbar v_{\rm D} \pi/a$, $a=3.5$~\AA, and $\epsilon_{\rm M}=1$.
\label{fig:qpole}}
\end{figure*}

Similarly to Sect.~\ref{sec:quasi}, we observe that for small $|\Im(\epsilon_\parallel)|\ll1$ there is a condition 
\begin{equation}\label{eq:qfrolich}
2\Re(\epsilon_\parallel) + m\epsilon_\bot + 3 \epsilon_{\rm M} = 0~,
\end{equation}
which determines the resonance excitation energies for quadrupole LSP resonances, at which an increased electromagnetic absorption occurs. 
If we take $\epsilon_\bot =0$, we recover the quadrupole Fr\"ohlich condition applicable to an isotropic conventional metallic sphere~\cite{hayashi_jpd_2012}, given by $2 \epsilon_{\parallel}+3 \epsilon_{\rm M}=0$. 
Fig.~\ref{fig:qpole}~a) illustrates the criteria for quadrupole LSP resonance, employing the same microscopic parameters as the dipole case depicted in Fig.~\ref{fig:dipole}. Unlike a conventional metal described by the Drude model~\cite{maier_book}, the resonant condition is fulfilled for five different energies instead of a single excitation energy.
Similarly to the dipole case, the axion term introduces a lifting of degeneracy in the isotropic conventional metallic sphere.
Therefore, when the resonance condition is satisfied for a given $m\in\{0,\pm1,\pm2\}$, the self-sustained electric field arises from the particular oscillating quadrupole moment $Q_{2m}$, defined in Eq.~\eqref{eq:Q2m}.
As discussed in Section \ref{sec:quasi}, we approximate the interband component of the optical conductivity as a renormalization of the dielectric constant $\epsilon_{\rm b}$, given by $\epsilon_{\rm b}\to \epsilon_{\rm b}^\star(w_{\rm Q})$, defined in Eq.~\eqref{eq:eps*},
%
%
where $w_{\rm Q}=\hbar \omega_{\rm pl}/\sqrt{\epsilon_{\rm b}+3\epsilon_{\rm M}/2}$ denotes the quadrupole LSP energy, excluding the axion term and interband optical conductivity.  
Thus, replacing $\epsilon_\parallel$ with $\epsilon_{\rm b}^\star(w_{\rm Q})-\frac{\omega_{\rm pl}^2}{\omega(\omega + i\gamma)}$ in Eq.~\eqref{eq:qfrolich}, and taking the ultraclean limit $\gamma=0^+$, enable us to obtain analytical expressions for the energies of plasmonic resonances easily
\begin{equation}\label{eq:qres}
\begin{aligned}
\Omega_{{\rm Q}m}&=\sqrt{\frac{\hbar^2\omega_{\rm pl}^2}{\epsilon_{\rm b}^\star(w_{\rm Q})+3 \epsilon_{\rm M}/2}+\bigg[\frac{m \alpha \hbar c b}{2\pi (\epsilon_{\rm b}^\star(w_{\rm Q})+3 \epsilon_{\rm M}/2)}\bigg]^2}\\
&-\frac{m \alpha \hbar c b}{2\pi (\epsilon_{\rm b}^\star(w_{\rm Q})+3 \epsilon_{\rm M}/2)}~,
\end{aligned}
\end{equation}
with $m\in \{0,\pm1,\pm2\}$.
For the fixed value $b=0.02\pi/a$,  the vertical dashed lines in Fig.~\ref{fig:qpole}~a) represent these energies, marking the conditions necessary for quadrupole surface resonance. They lie in an energy range where $\Im(\epsilon_\parallel)=0$, ensuring an infinite excitation lifetime.
Fig.~\ref{fig:qpole}~b) shows the plasmonic resonance energies, calculated numerically by solving Eq.~\eqref{eq:qfrolich} (depicted by circles), as a function of $b$.  Here, we compare the analytical expressions defined in Eq.~$\eqref{eq:qres}$ (shown as solid lines), which show a good agreement with numerical results. 
A notable difference between the analytical and numerical results appears only when the plasmonic resonant energies approach the threshold energy $2 \epsF$, indicated by the horizontal dashed gray line.

To conclude this Section, we examine the impact of a small spherical WSM particle under the influence of an oscillating electric field denoted as $\bm{\mathcal{E}}_{\rm inc}\approx ({\bm E}_0+{\bm E}_{\rm Q}) e^{-i \omega t}$, where ${\bm E}_0$ is uniform and oriented along the general direction $\bm{u}$ as expressed in Eq.~\eqref{eq:E0}, and $\bm{E}_{\rm Q}$ represents a field with linear spatial dependence as shown in Eq.~\eqref{eq:EQstatic}. Concurrently, the small sphere is also subjected to an oscillating TM field $\bm{\mathcal{B}}_{\rm inc}\approx ({\bm B}_0+{\bm B}_{\rm Q}) e^{-i \omega t}$, where ${\bm B}_0$ and ${\bm B}_{\rm Q}$ follow the definitions provided in Eqs.~\eqref{eq:B0} and~\eqref{eq:BQ}.
Examining up to order $(\sqrt{\epsilon_{\rm M}}\omega R/c)^2$, and employing the following property of the vector spherical harmonics:
\begin{equation}\label{eq:VHS}
\int d \Omega \bm{e}_r \cdot {\bm \Psi}_{\ell m}(\theta,\phi) \times {\bm \Phi}_{\ell' m'}^*(\theta,\phi) = \ell(\ell+1) \delta_{\ell \ell'} \delta_{m m'}~,
\end{equation}
where $\int d \Omega = \int_0^{2\pi} d\phi \int_0^\pi d \theta \sin(\theta)$, we obtain that the total energy absorption rate consists of a sum of two separate components $W_{\rm tot} \approx W_{0-\rm ext}+W_{\rm Q -ext}$, defined in Eqs.~\eqref{eq:Wext} and \eqref{eq:WQext}. In particular, the absorption rate consists of two independent components.

\section{A WSM sphere in a transverse electric field with a linear spatial dependence}
\label{sec:TE}

In this Section, within the quasistatic approximation, we study the interaction between a small spherical WSM particle and the oscillating and spatially uniform magnetic field $\bm{\mathcal{B}}_{\cal Q}= \bm{B}_{\cal Q} e^{-i \omega t}$, where
\begin{equation}\label{eq:TEBQ}
\bm{B}_{\cal Q}=\sum_{m=-1}^1 {\cal B}_m [\bm{Y}_{1m}(\theta,\phi)+  \bm{\Psi}_{1m}(\theta,\phi)]~,
\end{equation}
and the associated oscillating transverse electric (TE) field is $\bm{\mathcal{E}}_{\cal Q}= \bm{E}_{\cal Q} e^{-i \omega t}$, with
\begin{equation}\label{eq:TEQ}
\bm{E}_{\cal Q}=-i \frac{\omega}{2c}\sum_{m=-1}^1{\cal B}_m r \bm{\Phi}_{1m}(\theta,\phi)~,
\end{equation}
such that $\bm{B}_{\cal Q}=-i (c/\omega)\nabla\times \bm{E}_{\cal Q}$.
Inside the small sphere, such that $\omega R/c \ll 1$, the scattered electric field generated is expressed as $\bm{\mathcal{E}}_{{\cal Q}-\rm sc-in}\approx  \bm{E}_{\rm in}e^{-i \omega t}$, which we decompose as $\bm{E}_{\rm in}=\bm{E}_{\rm in}'+\bm{E}_{\rm in}'' +\bm{E}_{\rm in}'''$, where each contribution is defined in Eq.~\eqref{eq:EQin}, while the corresponding terms of the electric displacement vector $\bm{D}_{\rm in}=\bm{D}_{\rm in}'+\bm{D}_{\rm in}''+\bm{D}_{\rm in}'''$ are shown in Eq.~\eqref{eq:DQin}.
The associated scattered magnetic field is $\bm{\mathcal{B}}_{{\cal Q}-\rm sc-in}\approx  \bm{B}_{\rm in}e^{-i \omega t}$,   where $\bm{B}_{\rm in}=\bm{B}_{\rm in}'+\bm{B}_{\rm in}''$, 
with
\begin{equation}
\bm{B}_{\rm in}'= i \frac{2c}{\omega}\sum_{m=-1}^1   X_m'[\bm{Y}_{1m}(\theta,\phi)+  \bm{\Psi}_{1m}(\theta,\phi)]~,
\end{equation}
such that $\bm{B}_{\rm in}'=-i\frac{c}{\omega} \nabla\times \bm{E}_{\rm in}'$, 
and $\bm{B}_{\rm in}''$ defined in Eq.~\eqref{eq:Binpp},  such that $\nabla \times {\bm B}_{\rm in}''=-i (\omega/c) \bm{D}_{\rm in}''$.
Similarly, outside the sphere, in the near-zone region, we express the scattered electric field as $\bm{\mathcal{E}}_{{\cal Q}-\rm sc-out}\approx\bm{E}_{\rm out}e^{-i \omega t}$, where $\bm{E}_{\rm out}=\bm{E}_{\rm out}'+\bm{E}_{\rm out}''$, all contributions are defined in Eq.~\eqref{eq:EQout}, and the electric displacement vector is proportional to the electric field as $\bm{D}_{\rm out}=\epsilon_{\rm M} \bm{E}_{\rm out}$.
The associated scattered magnetic field is $\bm{\mathcal{B}}_{{\cal Q}-\rm sc-out}\approx \bm{B}_{\rm out} e^{-i \omega t}$,
%
%
where $\bm{B}_{\rm out}=\bm{B}_{\rm out}'+\bm{B}_{\rm out}''$, with
\begin{equation}
\begin{aligned}
\bm{B}_{\rm out}'&=i \frac{c}{\omega}\sum_{m=-1}^1 
\frac{Z'_m}{r^3}   [2 \bm{Y}_{1m}(\theta,\phi)-  \bm{\Psi}_{1m}(\theta,\phi)]~,
\end{aligned}
\end{equation}
such as $\bm{B}_{\rm out}'=-i (c/\omega) \nabla\times \bm{E}_{\rm out}'$, 
and $\bm{B}_{\rm out}''$, defined in Eq.~\eqref{eq:Boutpp},
%
such that $\nabla \times {\bm B}_{\rm out}''=-i (\omega/c) \bm{D}_{\rm out}''$.
%
By enforcing the continuity of the magnetic field along the surface of the sphere, one has the following conditions 
\begin{subequations}\label{eq:BTM}
\begin{align}
X_m'&=-i\frac{\omega}{2 c}{\cal B}_m+\frac{Z'_m}{R^3}~,\label{eq:contBa}\\
X_m'&=-i\frac{\omega}{2 c}{\cal B}_m-\frac{Z'_m}{2R^3}~,\label{eq:contBb}\\
X_{Dm}''&=-\frac{3}{2} \epsilon_{\rm M}\frac{Z_m''}{R^3}~,\label{eq:contBc}
\end{align}
\end{subequations}
with $m\in\{0,\pm1\}$.
To ensure the continuity of the normal components of the electric displacement vector, as indicated in Eq.~\eqref{eq:bcD}, Eq.~\eqref{eq:contBc} needs to be resolved by imposing $X_D'''=0$, corresponding to Eq.~\eqref{eq:Xppp}. In addition, the continuity of the transverse components of the electric field, as shown in Eq.~\eqref{eq:bcE}, is maintained by using Eq.~\eqref{eq:contBa} and adding the condition below
\begin{equation}\label{eq:ETM}
\begin{aligned}
X_m''&= \frac{Z_m''}{R^3}~.
\end{aligned}
\end{equation}
All boundary conditions are satisfied by the following parameters
\begin{subequations}
\begin{align}
X_m'&=-i \frac{\omega}{2 c} {\cal B}_m~,\\
Z_m'&=0~,\\
X'''&=-\frac{ \epsilon_\bot }{ \epsilon_\parallel }\frac{  \omega}{\sqrt{3} c } {\cal B}_0~,\\
X_0''&=\frac{Z_0''}{R^3}=\frac{\epsilon_\bot }{2\epsilon_{\parallel}+3\epsilon_{\rm M}} \frac{\omega}{\sqrt{15}c} {\cal B}_0~,\label{eq:X0ppTE}\\
X_{\pm 1}''&=\frac{Z_{\pm 1}''}{R^3}=\frac{\epsilon_\bot }{2\epsilon_{\parallel}\pm \epsilon_\bot+3\epsilon_{\rm M}} \frac{\omega}{2 \sqrt{5}c} {\cal B}_{\pm 1}~. \label{eq:XmppTE}
\end{align}
\end{subequations}
%
%
Here, 
for $|\Im(\epsilon_\parallel)|\ll1$, under the quadrupole LSP condition expressed in Eq.~\eqref{eq:qfrolich} with $m\in\{0,\pm1\}$,  we find that an infinitesimally oscillating uniform TE can induce a localized quadrupole plasmon mode characterized by the electric field $\bm{E}_{\rm out}''$ (
cf.~ Eqs.~\eqref{eq:X0ppTE} and \eqref{eq:XmppTE}). 
This phenomenon is unachievable in a conventional isotropic metallic sphere ~\cite{akimov_book}. 
We clarify that the homogeneous TE cannot excite a quadrupole plasmon mode labeled by $m=\pm 2$.
Within the framework of the quasistatic approximation, although an LSP can be produced under the conditions of a TE field, all electromagnetic power contributions disappear below the order of $(\sqrt{\epsilon_{\rm M}}\omega R/c)^3$.
Subsequently, we delve further into the origin of LSPs stimulated by an incoming TE field.
Starting from the Gauss's law, Eq.~\eqref{eq:nablaE}, we can express the polarization vector as satisfying the condition $-\nabla \cdot \bm{P}=\rho + \frac{\alpha}{2 \pi^2} \bm{b}\cdot\bm{B}$. By isolating the contribution of the polarization vector due to the axion term, one has $\bm{P}_{\rm ax}=-\frac{\alpha}{2\pi^2} \bm{b}\cdot \bm{r} \bm{B}$, leading to the associated electric field given by $\bm{E}_{\rm ax}=-4\pi(\mathbb{1}-\epsilon)^{-1}\cdot \bm{P}_{\rm ax}$, where $\mathbb{1}$ represents a $3\times3$ identity matrix. Specifically, this is expressed as $\bm{E}_{\rm ax}=\frac{2 \alpha}{\pi} \bm{b}\cdot\bm{r}(\mathbb{1}-\epsilon)^{-1}\cdot \bm{B}$. This is not a purely transverse electric field, since the magnetic field associated with a TE field possesses a non-zero component normal to the surface of the sphere. Hence, the $\bm{E}_{\rm ax}$ contains a component that aligns with the magnetic field.
The self-sustained electric field associated with an LSP, induced by an incident TE mode, which exhibits a finite component normal to the surface of the sphere, is contained in $\bm{E}_{\rm ax}$. The origin of LSP induced by an incoming TE field is the axion magnetoelectric effect~\cite{sekine_jap_2021}.

We conclude by analyzing the effect of a small spherical WSM particle when exposed to an external electric field expressed as $\bm{\mathcal{E}}_{\rm inc}= ({\bm E}_0+{\bm E}_{\rm Q}+{\bm E}_{\cal Q}) e^{-i \omega t}$. Here, ${\bm E}_{\cal Q}$ is the oscillating TE field as outlined in Eq.~\eqref{eq:TEQ}. Added to this is a uniform electric field ${\bm E}_0=E_0 \bm{u}$, as specified in Eq.~\eqref{eq:E0}, and an electric field exhibiting linear spatial variation $\bm{E}_{\rm Q}$, detailed in Eq.~\eqref{eq:EQstatic}. Simultaneously, the corresponding magnetic field is expressed as $\bm{\mathcal{B}}_{\rm inc}= ({\bm B}_0+{\bm B}_{\rm Q}+{\bm B}_{\cal Q}) e^{-i \omega t}$, with each component defined in Eq.~\eqref{eq:B0}, \eqref{eq:BQ}, and \eqref{eq:TEBQ}, respectively.
When $\sqrt{\epsilon_{\rm M}}\omega R/c \ll 1$, absorption primarily governs the extinction. Up the order of $(\sqrt{\epsilon_{\rm M}}\omega R/c)^2$, the total absorption rate is given by $W_{\rm tot}\approx W_{\rm ext}+W_{\rm Q -ext}+ W_{\rm mix}$, where the first two contributions are defined in Eqs.~\eqref{eq:Wext} and \eqref{eq:WQext}. The third contribution is
\begin{equation}
W_{\rm mix}=- 
\frac{5\epsilon_{\rm M} \omega^2 R^4}{8 \pi c}
 \epsilon_\bot  \sum_{m=-1}^{1}
f_m \Im \Big(\frac{ {\cal B}_{{\cal Q}m} E_{{\rm Q}m}^*}{2 \epsilon_{{\rm Q}m} +3\epsilon_{\rm M}} \Big)~,
\end{equation} 
with $f_0=1/\sqrt{15}$ and $f_{\pm 1}=1/(2\sqrt{5})$. 
The last term $W_{\rm mix}=-\int d^3 \bm{r} \delta(r-R) \bm{e}_r\cdot {\bm S}_{\rm mix}$ is obtainted from the contribution of the Poynting vector 

\begin{equation}
{\bm S}_{\rm mix}= \frac{c}{8\pi} 
  \Re[
\bm{\mathcal{E}}_{\rm Q}   
\times  \bm{\mathcal{B}}_{{\cal Q}-\rm sc-out}^\ast +\bm{\mathcal{E}}_{{\cal Q}-\rm sc-out} \times  \bm{\mathcal{B}}_{\rm Q} ^\ast     ]~,
\end{equation}
it involves an incident electromagnetic field exhibiting quadrupole symmetry, characterized by a TM field, $\bm{\mathcal{E}}_{\rm Q}$ and $\bm{\mathcal{B}}_{\rm Q}$, and the electromagnetic field produced by the sphere scattering the incident electromagnetic field, also showing a quadrupole symmetry but characterized by a TE field, $\bm{\mathcal{E}}_{{\cal Q}-\rm sc-out}$ and $\bm{\mathcal{B}}_{{\cal Q}-\rm sc-out}$. This mixing term emerges directly from the axion component, a phenomenon not usually observed in conventional metals~\cite{akimov_book}. It introduces an additional energy absorption pathway that can be advantageous for thermoplasmonic applications.

\section{Summary and Conclusions}
\label{sec:conclusions}

In this work, we have theoretically investigated the LSP modes in a sub-wavelength spherical nanoparticle made of a WSM, accounting for the axion contribution to the electromagnetic response. Within the quasistatic approximation, we solved the modified Maxwell equations incorporating the $\theta$-term and characterized both the dipole and quadrupole plasmonic excitations. The dielectric response of the WSM is described by a non-diagonal, gyrotropic term, arising from the time-reversal symmetry-breaking term $\bm{b}$.
For dipole symmetric LSPs, we derived the modified Fr\"ohlich condition, which, unlike conventional isotropic metals, leads to three distinct resonance frequencies due to the axion term, $\epsilon_\bot$. These resonances correspond to a nontrivial polarization dependence of the absorption cross section, highlighting the anisotropic and topological nature of the electromagnetic response.
The analysis was extended to the quadrupole symmetric LSPs. In conventional isotropic metallic nanoparticles, there is typically one quadrupole resonance condition. Here, thanks to the axion term, this number rises to five. We derived analytical expressions for the resonant frequencies, illustrating their dependence on the axion term.
Finally, we examined the excitation of plasmonic modes by TE fields. Notably, the axion magnetoelectric effect enables the excitation of LSP, a phenomenon that does not occur in conventional, isotropic metallic nanospheres. Moreover, we identified a mixing term in the absorption cross-section due to interference between incident TE and TM fields, again arising from the axion term.
In conclusion, our results emphasize the intriguing and tunable plasmonic properties of WSM nanoparticles, directly stemming from their topological electromagnetic response. The modification of the Fr\"ohlich conditions due to the axion term leads to richer LSP resonance spectra and polarization dependencies. 
Since the WSM phase is robust against weak disorder~\cite{buchhold_prb_2018}, it is reasonable to infer that the described splitting in the dipole and quadruple resonances shares such robustness. 
These results could pave the way for applications of WSM nanostructures in anisotropic nanoplasmonics, nonreciprocal photonics, and potentially in chiral sensing or topological thermoplasmonic technologies. 

\begin{acknowledgments}
The authors thank A. Biondo, A. Cupolillo, E. Curcio, A. Politano, and B. Tomasello for illuminating discussions and fruitful comments on various stages of this work. F.M.D.P acknowledges support from the project PRIN 2022 - 2022XK5CPX (PE3) SoS-QuBa - "Solid State Quantum Batteries: Characterization and Optimization" and the PNRR MUR project PE0000023-NQSTI.
G.G.N.A. acknowledges support from the project MUR PRIN PNRR - P20223LXTA - ENTANGLE. F.B. acknowledges financial support from the TOPMASQ Project, CUP E13C24001560001, funded by PE0000023-NQSTI.
\end{acknowledgments}

\bibliography{literaturew}%

\end{document}